\documentclass[prb,aps,floatfix]{revtex4}
\usepackage{amsmath}
\usepackage{amssymb}
\usepackage{amsthm}
\usepackage{epsfig}
\usepackage{tikz}
\usepackage{enumitem}
\usepackage{etoolbox}
\usepackage{xcolor}
\def\d{ {\mathrm{d}}}
\def\up{\uparrow}
\def\pd{ {\phantom\dagger} }
\def\down{\downarrow}
\def\nn{\nonumber\\}
\def\fa#1#2{{a_#1(#2)}}
\def\l{\Lambda}

\newcommand{\be}{\begin{equation}}
\newcommand{\ee}{\end{equation}}
\newcommand{\beA}{\begin{equation}\begin{aligned}}
\newcommand{\eeA}{\end{aligned}\end{equation}}
\newcommand{\bem}{\begin{multline}}
\newcommand{\eem}{\end{multline}}
\newcommand{\bea}{\begin{eqnarray}}
\newcommand{\eea}{\end{eqnarray}}
\theoremstyle{definition}

\theoremstyle{remark}

\def\fr#1{(\ref{#1})}

\begin{document}
\title{Atypical energy eigenstates in the Hubbard chain and quantum
  disentangled liquids}  
\author{Thomas Veness}
\affiliation{The Rudolf Peierls Centre for Theoretical Physics, University of
Oxford, Oxford OX1 3NP, UK}

\author{Fabian H.~L.~Essler}
\affiliation{The Rudolf Peierls Centre for Theoretical Physics,
  University of Oxford, Oxford OX1 3NP, UK}

\author{Matthew P.A. Fisher}
\affiliation{Department of Physics, University of California, Santa
Barbara, California 93106, USA}

\begin{abstract}
We investigate the implications of integrability for the existence of quantum
disentangled liquid (QDL) states in the half-filled one-dimensional Hubbard
model. We argue that there exist finite energy-density eigenstates that exhibit
QDL behaviour in the sense of J. Stat. Mech. P10010 (2014).
These states are atypical in the sense that their entropy density is smaller
than that of thermal states at the same energy density. Furthermore,
we show that thermal states in a particular temperature window exhibit
a weaker form of the QDL property, in agreement with recent results
obtained by strong-coupling expansion methods in arXiv:1611.02075. 
This article is part of the themed issue `Breakdown of ergodicity in quantum
	systems: from solids to synthetics matter'.
\end{abstract}

\maketitle
\section{Introduction}
The question of how isolated many-particle quantum systems relax and
how to describe their steady state behaviour has attracted attention
for a long time\cite{vonNeumann}. The past decade has witnessed a
tremendous resurgence of interest in this problem, which was largely
motivated by ground-breaking experiments on systems of trapped
ultra-cold
atoms\cite{GM:col_rev02,kww-06,HL:Bose07,hacker10,tetal-11,getal-11,cetal-12,langen13,MM:Ising13,zoran1,MBLex1}. It
is now understood that generic many-body systems relax towards thermal
equilibrium distributions at an effective temperature fixed by the
energy density, which is, by definition, conserved for isolated
systems. This behaviour follows from the eigenstate thermalisation
hypothesis (ETH)\cite{Deutsch91,Sred1,Sred2,ETH}. When a system
thermalises the only information about the initial state that is
retained at late times is its energy density. 
This does not, however, exhaust the theoretically understood paradigms of
relaxation: quantum integrable systems possess conservation laws that constrain
the system to retain information on more than just the energy density.  
As such, they do not thermalise, but instead relax towards Generalised Gibbs
Ensembles \cite{GGE,cc-07,EFreview,Cazalilla,VR,GGE_int}.
This can be understood in terms of a generalised ETH\cite{GETH,CE13}.
Sufficiently strong disorder is another mechanism that can preclude
thermalisation\cite{MBL1,MBL2,MBL3,MBL4,MBL5,MBL6}. This can again 
be related to the existence of conservation laws\cite{MBL6,MBL7,MBL8},
although, unlike in the integrable case, no fine-tuning is
required. Moreover, in (many-body) localised systems eigenstates at
finite energy densities exhibit an area-law scaling of the
entanglement entropy (EE). This is qualitatively different to cases in
which the ETH holds and differs dramatically from the situation
encountered in integrable models.

Recently it has been proposed that the eigenstates of certain systems may
fail to thermalise in the conventional sense. The corresponding state
of matter has been dubbed the ``quantum disentangled liquid'' (QDL)\cite{qdl}. 
A characteristic feature of such systems is that they comprise both heavy and
light degrees of freedom. The basic premise of the QDL concept is that,
while the heavy degrees of freedom are fully thermalised, the light
ones, which are enslaved to the heavy particles, are not
independently thermalised.  A convenient diagnostic for such a state
of matter is the bipartite EE after a
projective measurement of the heavy particles. 
The possibility of realising a QDL in the one-dimensional Hubbard model was
subsequently investigated by exact diagonalisation of small systems in
Ref.~\onlinecite{hubb1}. Given the limitations on accessible system sizes it is
difficult to draw definite conclusions from these results. Motivated by these
studies we have recently explored the possibility of realising a QDL
in the half-filled Hubbard model on bipartite lattices by analytic
means\cite{QDLHubbard}. Here we provide additional details regarding
the integrability-based approach put forward in that work. The
Hamiltonian of the one-dimensional Hubbard model is
\be
H=-t\sum_{j,\sigma=\up,\down}
\left(
c^\dagger_{j,\sigma}c^\pd_{j+1,\sigma}
+c^\dagger_{j+1,\sigma}c^\pd_{j,\sigma}
\right)
+U\sum_{j}
\big(n_{j,\uparrow}-\frac{1}{2}\big)
\big(n_{j,\downarrow}-\frac{1}{2}\big).
\label{HHubb}
\ee
Here $c^\pd_{j,\sigma}$, $c^\dagger_{j,\sigma}$ are fermionic operators
satisfying the usual anticommutation relations,
$n_{j,\sigma}=c^\dagger_{j,\sigma}c^\pd_{j,\sigma}$, $t>0$ is the hopping
parameter and $U>0$ is the strength of the on-site repulsion.

The outline of this paper is as follows. In
Section~\ref{sec:integrability} we briefly recall necessary facts from
the exact solution of the Hubbard model.
In Section~\ref{sec:typical} we consider typical states at finite temperature.
In Sections~\ref{sec:Heis}and \ref{sec:five} we employ methods of integrability to show
that it is possible to construct particular eigenstates at finite energy
densities for which the charge degrees of freedom do not contribute to the
volume term in the bipartite EE, corroborating the notion of the QDL diagnostic
proposed in Ref.~\onlinecite{qdl}. 
In Section~\ref{sec:thermal} we show that there exists a parametrically large
regime in which thermal states support a weaker version of QDL as proposed in
Ref.~\onlinecite{QDLHubbard}.
\section{Eigenstates of the Hubbard Hamiltonian}
\label{sec:integrability}
The Bethe Ansatz method provides an exact solution of the one-dimensional
Hubbard model\cite{book}. Within the framework of the string
hypothesis, eigenstates in the Hubbard model are determined by
solutions to \emph{Takahashi's equations}\cite{book}. For a state with
$N$ electrons, $M$ of which are spin-down, they read  
\beA
\frac{2\pi I_j}{L} &= k_j +  \frac{1}{L} \sum_{n=1}^\infty \sum_{\alpha=1}^{M_n} \theta \left( \frac{\sin k_j - \Lambda^n_\alpha}{nu}\right) 
+ \frac{1}{L} \sum_{n=1}^\infty \sum_{\alpha=1}^{M_n'}\theta\left( \frac{\sin k_j - {\Lambda'}_\alpha^n}{nu}\right)  ,&&\qquad j=1,\dots, N-2M', \\
\frac{2\pi J^n_\alpha}{L}
&= \frac{1}{L}\sum_{j=1}^{N-2M'} \theta\left( \frac{\Lambda^n_\alpha - \sin k_j}{nu}\right) - \frac{1}{L}\sum_{m=1}^\infty \sum_{\beta=1}^{M_m} \Theta_{nm} \left( \frac{\Lambda^n_\alpha - \Lambda^m_\beta}{u} \right),&&\qquad \alpha=1,\dots,M_n,\\
\frac{2\pi {J'}^n_\alpha}{L}&= -\frac{1}{L}\sum_{j=1}^{N-2M'} \theta\left( \frac{ {\Lambda'}^n_\alpha - \sin k_j}{nu}\right) - \frac{1}{L}\sum_{m=1}^\infty \sum_{\beta=1}^{M'_m} \Theta_{nm} \left( \frac{ {\Lambda'}^n_\alpha - {\Lambda'}^m_\beta}{u} \right) \nn
&+ 2{\rm Re}[\arcsin( {\Lambda'}^n_\alpha + niu)],&&\qquad \alpha=1,\dots,M_n',
\label{BAE}
\eeA
where $u=U/4t$, $\theta(x) = 2 \arctan(x)$,  
\be
\Theta_{nm}(x) = \begin{cases} \theta\left( \frac{x}{|n-m|} \right) + 2\theta\left( \frac{x}{|n-m|+2} \right)+ \dots + 2\theta\left( 
\frac{x}{n+m-2} \right) + \theta\left( \frac{x}{n+m} \right),& n\neq m\\
2\theta\left( \frac{x}{2} \right) + 2\theta\left( \frac{x}{4} \right) + \dots + 2\theta\left( \frac{x}{2n-2} \right) + \theta\left( \frac{x}{2n} \right),& n=m \end{cases},
\ee
and
\be
M = \sum_{n=1}^\infty n(M_n + M_n'), \qquad M' = \sum_{n=1}^\infty n M_n'.
\ee
The sets $\{ I_j\}$, $\{ J^n_\alpha\}$, $\{ {J'}^n_\alpha \}$ of
integer or half-odd integer numbers specify the particular
eigenstate under consideration and obey the ``selection rules''
\be
\begin{aligned}
I_j &\in \begin{cases} \mathbb{Z} + \frac{1}{2} & {\rm if}\ \sum_m (M_m +M'_m)\ {\rm odd}\\
       \mathbb{Z} & {\rm if}\ \sum_m (M_m +M'_m)\ {\rm even}
\end{cases},&& \qquad -\frac{L}{2} < I_j \leq \frac{L}{2},\\
J^n_\alpha &\in \begin{cases} \mathbb{Z} & {\rm if}\ N-M_n\ {\rm odd}\\
       \mathbb{Z}+\frac{1}{2} & {\rm if}\ N-M_n\ {\rm even}
\end{cases},&& \qquad |J^n_\alpha| \leq \frac{1}{2} (N-2M' - \sum_{m=1}^\infty t_{nm} M_m -1),\\
{J'}^n_\alpha &\in \begin{cases} \mathbb{Z} & {\rm if}\ L-N+M'_n\ {\rm odd}\\
       \mathbb{Z}+\frac{1}{2} & {\rm if}\ L-N+M'_n\ {\rm even}
\end{cases},&&
\qquad |J'^n_\alpha| \leq \frac{1}{2}\left( L - N + 2M' - \sum_{m=1}^\infty t_{nm} M_m' - 1 \right),
\label{eq:integerParity}
\end{aligned}
\ee
where $t_{nm}= 2\, { \rm min} (m,n) - \delta_{mn}$. Energy and
momentum, measured in units of $t=1$, of an eigenstate characterised
by the set of roots $\{ k_j, \Lambda_\alpha^n, {\Lambda'}_\beta^m\}$ are given by 
\be
E = - 2\sum_{j=1}^{N-2M'} \cos k_j + + 4 \sum_{n=1}^\infty \sum_{\beta=1}^{M_n'} {\rm Re} \sqrt{ 1 - ( {\Lambda'}^n_\beta + niu)^2} + u(L-2N), \label{eq:energyFull}
\ee
\be
       P = \left[  \sum_{j=1}^{N-2M'} k_j - \sum_{n=1}^\infty \sum_{\beta=1}^{M_n'} \left( 2\,{\rm Re} \arcsin\left( {\Lambda'}^n_\beta + niu \right) - (n+1)\pi \right) \right] {\rm mod}\, 2\pi.
\ee
In the framework of the string hypothesis each set $\{ I_j,
J_\alpha^n, {J'}_\beta^m \}$ of (half-odd) integers gives rise to a
unique eigenstate of the Hubbard Hamiltonian. In particular, the
ground state for even lattice length $L$, even total number of electrons
$N_{GS}$ and odd number of down spins $M_{GS}$ is obtained by the
choice\cite{book}  
\begin{align}
I_j &= - \frac{N_{GS}}{2} - \frac{1}{2} + j,&&  j=1,\ldots,N_{GS} ,\\
J^1_\alpha &= - \frac{M_{GS}}{2} - \frac{1}{2} + \alpha,&&  \alpha = 1,\ldots, M_{GS}. \label{eq:GSints}
\end{align}

\subsection{Macro states at finite energy densities}
Taking the thermodynamic limit, the Bethe Ansatz allows a description
of macro states corresponding to smooth root distributions. We now
use this framework to identify a class of macro states that exhibits 
characteristic properties of a QDL. Using the string hypothesis,
general macro states in the one-dimensional Hubbard model can be
described by sets of particle and hole densities 
$\{\rho^{p}(k), \rho^{h}(k),
\sigma^{p}_n(\Lambda), \sigma^{h}_n(\Lambda),
{\sigma'_n}^{p}(\Lambda), {\sigma'_n}^{h}(\Lambda)
|n\in\mathbb{N}\}$ 
that are subject to the thermodynamic limit of the Bethe Ansatz
equations\cite{book}
\begin{widetext}
\index{root density}
\begin{eqnarray}
&&\rho^p(k)+\rho^h(k)=\frac{1}{2\pi}+\cos
k\sum_{n=1}^\infty\int_{-\infty}^\infty d\Lambda\ \fa{n}{\l-\sin k}
\left[{\sigma^\prime_n}^p(\Lambda)+\sigma_n^p(\Lambda)\right]\ ,\nn
&&\sigma_n^h(\Lambda)=-\sum_{m=1}^\infty
\int_{-\infty}^\infty d\Lambda' A_{nm}(\Lambda-\Lambda')\ \sigma_m^p(\Lambda')
+\int_{-\pi}^\pi dk\ \fa{n}{\sin k-\Lambda}\ \rho^p(k)\ ,\nn
&&{\sigma^\prime_n}^h(\Lambda)=\frac{1}{\pi}{\rm
Re}\frac{1}{\sqrt{1-(\Lambda -inu)^2}}
-\sum_{m=1}^\infty 
\int_{-\infty}^\infty d\Lambda' A_{nm}(\Lambda-\Lambda')\ {\sigma'_m}^p(\Lambda')
-\int_{-\pi}^\pi dk\ \fa{n}{\sin k-\Lambda}\ \rho^p(k)\ ,
\label{densities}
\end{eqnarray}
\end{widetext}
where we are considering expectation values only to
$o(1)$ and subleading corrections require a more detailed analysis.
These equations are obeyed for all macro states and are a direct implication of
the quantisation conditions in the thermodynamic limit.
Above, $u=U/4t$ and
\beA
a_n(x)&=\frac{1}{2\pi}\frac{2nu}{(nu)^2+x^2}\ ,\\
A_{nm}(x)&=\delta(x)+ (1-\delta_{m,n})a_{|n-m|}(x)+2a_{|n-m|+2}(x)
+\dots+2a_{|n+m|-2}(x)+a_{n+m}(x).
\label{Anm}
\eeA
The energy and thermodynamic entropy per site are then given by
\beA
e&=u+
\int_{-\pi}^\pi dk\left[-2\cos k -2u\right]\rho^p(k)
+ 4 \sum_{n=1}^\infty \int d\Lambda\,{\sigma^\prime_n}^p(\Lambda)\,\left[{\rm Re}\sqrt{1-(\Lambda+inu)^2} 
-4nu\right]
,\\
s&=\int_{-\pi}^\pi dk\ {\cal S}\left[\rho^p(k),\rho^h(k)\right]+u
+\sum_{n=1}^\infty\int_{-\infty}^\infty d\Lambda\
{\cal S}\left[{\sigma^\prime_n}^p(\Lambda),{\sigma_n^\prime}^h(\Lambda)\right]\\
&+\sum_{n=1}^\infty\int_{-\infty}^\infty d\Lambda\
{\cal S}\left[{\sigma_n}^p(\Lambda),{\sigma_n}^h(\Lambda)\right],
\label{entropy}
\eeA
where we have defined
\be
{\cal S}[f,g]=
\big[f(x)+g(x)\big]\ln\big(f(x)+g(x)\big)
-f(x)\ln\big(f(x)\big)
-g(x)\ln\big(g(x)\big)\ .
\ee
The ground state of the half-filled Hubbard model in zero magnetic field is
obtained by choosing
\be
\rho^h(k)=0=\sigma^h_1(\Lambda)\ ,\qquad
{\sigma'}_n^p(\Lambda)=0=
{\sigma}_{n\geq 2}^p(\Lambda)\ .
\ee
\section{Typical vs atypical energy eigenstates}
\label{sec:typical}
A characteristic property of integrable models is that, at finite
energy densities relative to the ground state, there exist thermal
states as well as \emph{atypical finite entropy density states} that
have rather different properties. The existence of such states is
intimately related to the presence of an extensive number of higher
conservation laws. Their nature can be easily understood by
considering the special limit of non-interacting fermions
($U=0$). Here the half-filled ground state is simply
\be
|{\rm  GS}\rangle_{U=0}=\prod_{\sigma,|k_j|<\pi/2}c^\dagger_{\sigma}(k_j)|0\rangle\ ,\quad
\ee
where $c_\sigma(k)=L^{-1/2}\sum_je^{ikj}c_{j,\sigma}$. Thermal states
at finite energy densities are Fock states with momentum distribution
function 
\be
\rho^p_\sigma(k)=\frac{1}{2\pi[1+e^{-2\cos(k)/T}]}\ .
\ee
In a large, finite volume, we can construct thermal Fock states by
using the relation
$\rho^p_\sigma(k_{j})=\frac{1}{L(k_{j+1}-k_j)}+o(1)$. A simple
atypical state at a finite energy density above the ground state is
obtained by splitting the Fermi sea 
\be
|{\rm split\ FS}\rangle=\prod_{\sigma,\frac{\pi}{4}<|k_j|<\frac{3\pi}{4}}c^\dagger_{\sigma}(k_j)|0\rangle\ .
\label{SFS}
\ee
The energy eigenstate \fr{SFS} is clearly not thermal. Moreover, the
corresponding macro state has zero entropy density in the
thermodynamic limit. However, it is easy to see that, by considering
other arrangements of the momentum quantum numbers, one can arrive at
atypical states that have finite entropy densities in the
thermodynamic limit \cite{ACF}. The situation in integrable models is a
straightforward generalisation of this construction. The relevant
quantum numbers are the (half-odd) integer numbers that characterise
the solutions of the Bethe Ansatz equations.

\subsection{Thermal states in the Hubbard model}
\label{app:TBAeqns}
Thermal states are, by construction, the most likely states at a given
energy density. To obtain their description in terms of particle and
hole distribution functions we need to maximise the entropy density
$s$ at a fixed energy density $e$. To that end, it is customary to extremise the
free energy per site $f=e-Ts$, where $e$ and $s$ are given in
\fr{entropy}
\bea
0=\delta f &=&\int_{-\pi}^\pi dk\left[
\frac{\delta f}{\delta\rho^p(k)}\delta\rho^p(k) 
+\frac{\delta f}{\delta\rho^h(k)}\delta\rho^h(k) \right]\nn
&&+\sum_{n=1}^\infty\int_{-\infty}^\infty d\Lambda
\left[\frac{\delta f}{\delta{\sigma^\prime_n}^p(\Lambda)}
\delta{\sigma^\prime_n}^p(\Lambda)
+\frac{\delta f}{\delta{\sigma^\prime_n}^h(\Lambda)}
\delta{\sigma^\prime_n}^h(\Lambda)
+\frac{\delta f}{\delta\sigma_n^p(\Lambda)}
\delta\sigma_n^p(\Lambda)
+\frac{\delta f}{\delta\sigma_n^h(\Lambda)}
\delta\sigma_n^h(\Lambda)\right].
\label{deltaf}
\end{eqnarray}
The relations \fr{densities} connect hole and particle densities and
need to be taken into account as constraints. The extremisation leads to
a system of non-linear integral equations that fixes the ratios
\be
\zeta(k)=\frac{\rho^h(k)}{\rho^p(k)}\ ,\quad
\eta_n(\Lambda)=\frac{\sigma^h_n(\Lambda)}{\sigma^p_n(\Lambda)}\ ,\quad
\eta'_n(\Lambda)=\frac{{\sigma'}^h_n(\Lambda)}{{\sigma'}^p_n(\Lambda)}.
\ee
For the Hubbard model in zero magnetic field the resulting
\emph{Thermodynamic Bethe Ansatz equations} read \cite{takahashiTBA1,book}
\begin{widetext}
\begin{eqnarray}
\ln \zeta(k)&=&\frac{-2\cos k -2u}{T}
+\sum_{n=1}^\infty 
\int_{-\infty}^\infty 
d\Lambda\ \fa{n}{\sin k-\Lambda}
\ln\left(1+\frac{1}{\eta^\prime_n(\Lambda)}\right)\nn
&-&\sum_{n=1}^\infty \int_{-\infty}^\infty
d\Lambda\ \fa{n}{\sin k-\Lambda}
\ln\left(1+\frac{1}{\eta_n(\Lambda)}\right),\nn
\ln\left(1+\eta_n(\Lambda)\right) &=& -\int_{-\pi}^\pi dk 
\ \cos(k)\ \fa{n}{\sin k-\Lambda}
\ln\left(1+\frac{1}{\zeta(k)}\right)
+\sum_{m=1}^\infty
A_{nm}*\ln\left(1+\frac{1}{\eta_m}\right)\bigg|_\Lambda\ ,\nn
\ln\left(1+\eta^\prime_n(\Lambda)\right)&=&
\frac{4{\rm Re}\sqrt{1-(\Lambda -inu)^2}-4nu}{T}
-\int_{-\pi}^\pi dk \ \cos(k)\ \fa{n}{\sin k-\Lambda}
\ln\left(1+\frac{1}{\zeta(k)}\right)\nn
&+&\sum_{m=1}^\infty A_{nm}*\ln\left(1+\frac{1}{\eta^\prime_m}\right)
\bigg|_\Lambda .
\label{TBAeqns}
\end{eqnarray}
\end{widetext}
The system \fr{TBAeqns} can be solved numerically to
calculate the energy density and other simple thermodynamic properties
of typical states at finite energy density.
The free energy per site is given in terms of the solution of (\ref{TBAeqns}) by\cite{book}
\be
f=-T\int_{-\pi}^\pi \frac{\d k}{2\pi} \ln \left( 1 + \frac{1}{\zeta(k)}\right) + u
-T\sum_{n=1}^\infty \int_{-\infty}^\infty \frac{\d \Lambda}{\pi} \ln\left(1+\frac{1}{\eta'_n(\Lambda)}\right) {\rm Re}\frac{1}{\sqrt{1-(\Lambda-inu)^2}}.
\label{eq:TBAf}
\ee

\subsection{Simple families of atypical finite entropy density states
  in the Hubbard model} 
\label{app:exactFamilies}
It is instructive to explicitly construct families of atypical macro
states with finite entropy densities, which allow one to obtain
closed-form expressions for the energy density and double occupancy
\be
d=\frac{1}{L} \Big\langle \sum_j n_{j,\up} n_{j,\down}\Big\rangle\ ,
\ee
in the thermodynamic limit. In terms of the Bethe Ansatz, the states
we wish to consider involve ``freezing'' the microscopic configuration
of the charge sector to that of the ground state at half-filling. More
precisely, we consider the following two-parameter family of macro states
\be
{\sigma'}^p_n(\Lambda)=0\ ,\quad \rho^h(k)=0\ ,\quad
\sigma_1^h(\Lambda) = x \sigma_1^p(\Lambda),\qquad \sigma_n^h(\Lambda) = y
\sigma_n^p(\Lambda).
\label{exactfamily}
\ee
The choice \fr{exactfamily} enables us to solve the thermodynamic
limit of the Bethe Ansatz equations \fr{densities} by Fourier
techniques. In particular we find that the Fourier transforms of the
particle densities in the spin sector
$\widetilde{\sigma}_n(\omega) = \int \d \Lambda e^{i \omega \Lambda}
\sigma^p_n(\Lambda)$ fulfil
\be
\begin{pmatrix} 1+ x + e^{-2 u |\omega} & e^{-(n-1)u|\omega|} +
	e^{-(n+1)u|\omega|} \\ e^{-(n-1)u|\omega|} + e^{-(n+1)u|\omega|} & 1 + y
	+ 2e^{-2u|\omega|} +\cdots + 2e^{-2(n-1)u|\omega|} + e^{2nu|\omega|}
\end{pmatrix} \begin{pmatrix} \widetilde{\sigma}_1(\omega) \\
	\widetilde{\sigma}_n(\omega)\end{pmatrix} = \begin{pmatrix} J_0(\omega)
	e^{-u|\omega|} \\ J_0(\omega) e^{-n u |\omega|} \end{pmatrix} ,
\ee
where $J_n(\omega)$ are Bessel functions of the first kind. Taking the
$\omega\to0$ limit, this gives 
\be
\begin{pmatrix} 2+x & 2 \\ 2 & y+2n\end{pmatrix}
\begin{pmatrix}\widetilde{\sigma}_1(0) \\ \widetilde{\sigma}_n(0) \end{pmatrix}
= \begin{pmatrix} 1 \\ 1 \end{pmatrix}.
\label{eq:matrixEq}
\ee
We are particularly interested in spin singlet states. By the theorem
of Refs~\onlinecite{HWS,HWS2} a sufficient condition for obtaining a
singlet is for the $S^z$ eigenvalue to be zero, which imposes the constraint
\be
\widetilde{\sigma}_1(0) + n\widetilde{\sigma}_n(0) = \frac{1}{2}.
\ee
Combining this with (\ref{eq:matrixEq}) leads to the $n$-independent requirement
$xy=0$. As $x=0$ corresponds to the ground state we choose $y=0$. This
corresponds to a finite density of holes for 1-strings and a filled
Fermi sea for $n$-strings. The energy density for the atypical macro
states constructed in this way is
\be
e_n(x) =-4\int_0^\infty \frac{\d \omega}{\omega}
J_0(\omega) J_1(\omega)
\frac{
1 + e^{-2 u \omega} + 
   e^{(2 - 2 n) u \omega} (-1 + x) - 
   e^{-2 n u \omega} (1 + x)  }{
  (1 + e^{-2 u \omega}) (1 - 
   e^{(4 - 2 n) u \omega} + e^{2 u \omega} (1 + x) - 
   e^{(2 - 2 n) u \omega} (1 + x)) 
}.
\label{enx}
\ee
\begin{figure}[ht]
\includegraphics[width=0.45\linewidth]{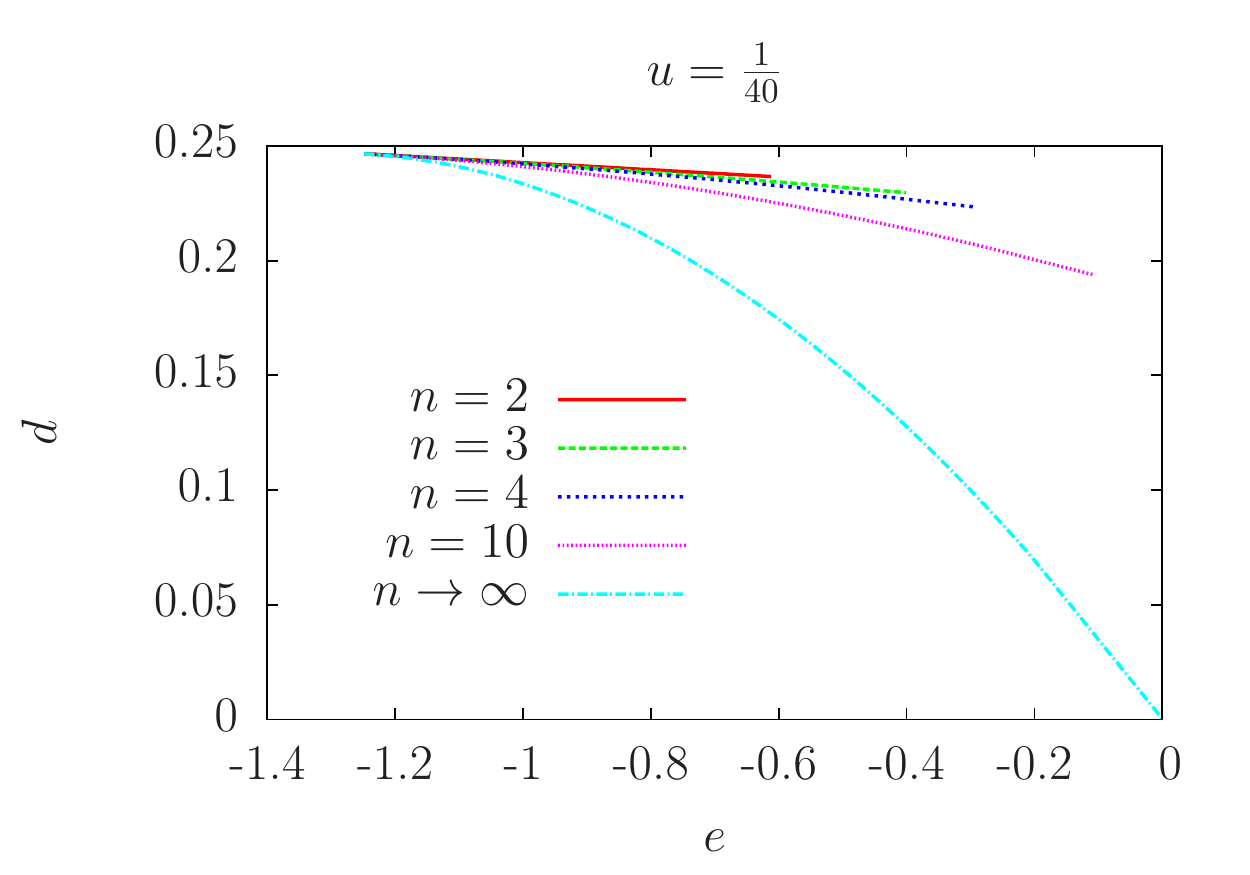}
\includegraphics[width=0.45\linewidth]{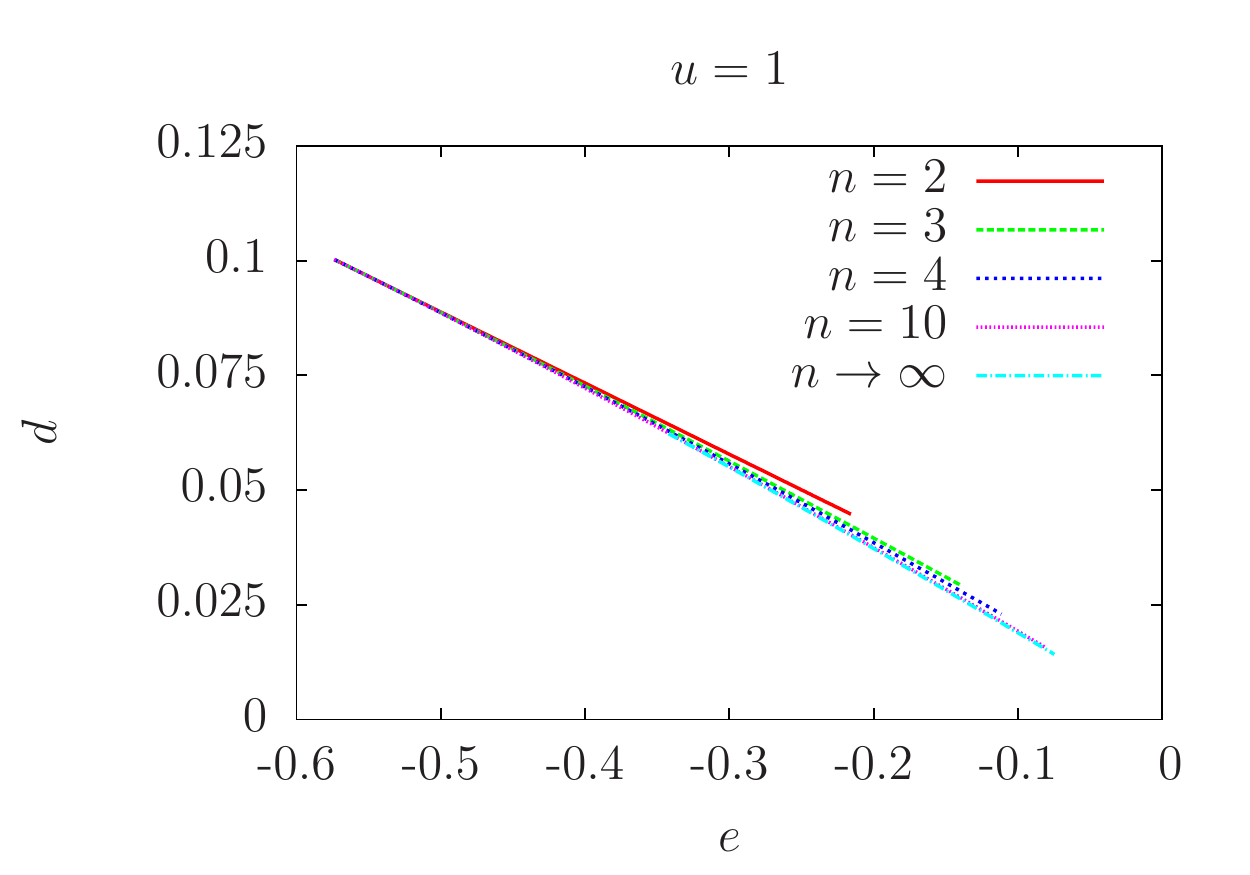}
\caption{$e$ vs $d$ curves for $u=1/40$ and $u=1$ respectively, showing states occupying 1-strings and
$n$-strings} \label{fig:curves}
\end{figure}
By taking derivatives with respect to $u$, we can calculate the double
occupancy $d$ as a function of $x$, and combining this with \fr{enx}
we can study how $d$ changes with $e$ for the different families $n$.
The results are shown in Fig.~\ref{fig:curves}.

\subsection{Double occupancy for thermal vs atypical states}
It is interesting to compare the behaviour of the double occupancy in thermal
states and the particular family of atypical states as identified above in
Section~\ref{app:exactFamilies}.
We can calculate the energy density and double occupancy for typical states
using the free energy of (\ref{eq:TBAf}) as
\be
\langle e\rangle_\beta = \frac{\partial }{\partial \beta}\big(\beta f\big),\qquad
\langle d\rangle_\beta = \frac{\partial f}{\partial U} +\frac{1}{4},
\ee
where we have used the fact that we are working at half-filling.
This determines $d$ as an implicit function of $e$ for thermal states.
We note that this is of experimental relevance, as recent ultra-cold
atomic experiments are able to directly measure the double occupancy
in realisations of the one-dimensional Hubbard model\cite{Bloch}.

\begin{figure}[ht]
\begin{center}
\includegraphics[width=0.8\linewidth]{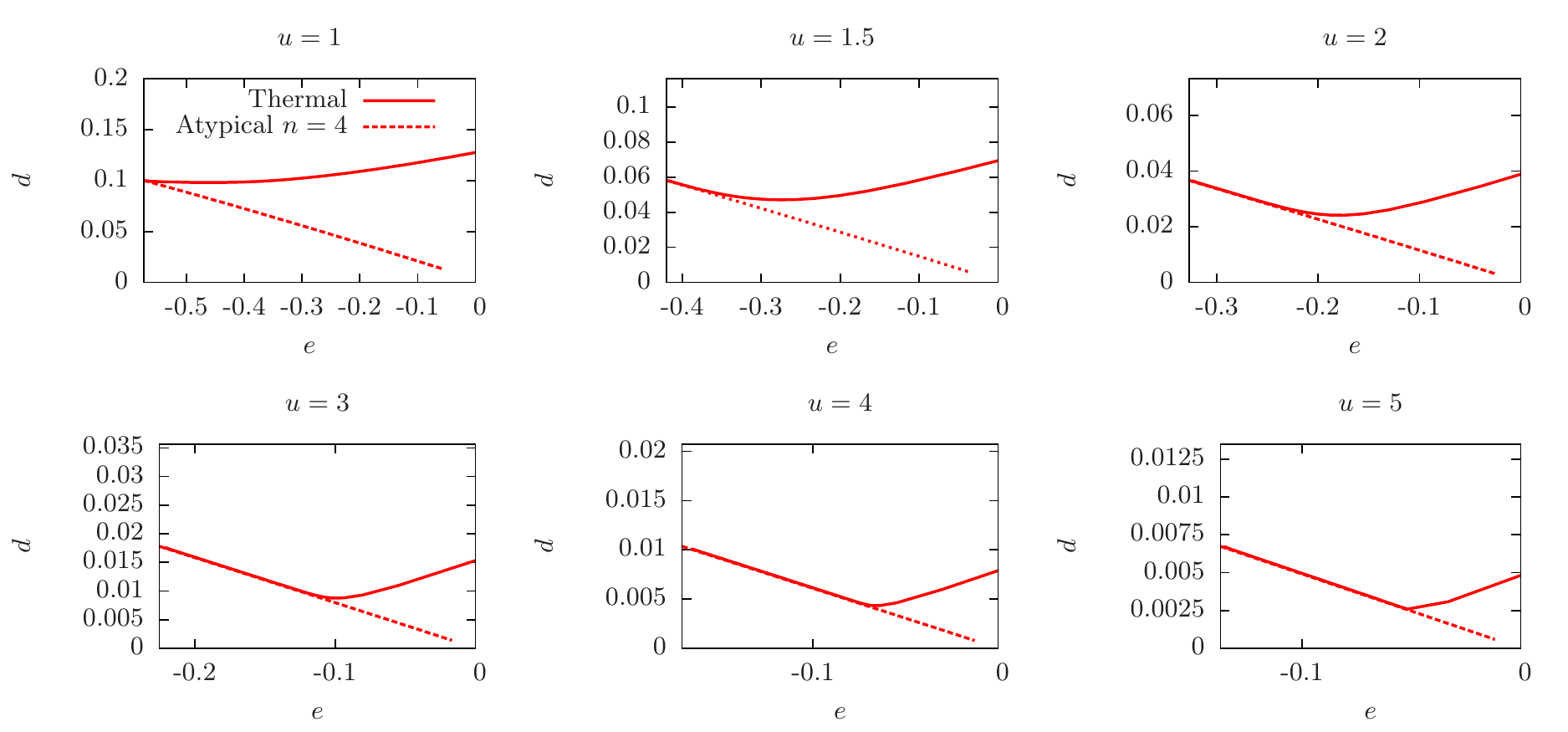}
\end{center}
\caption{Double occupancy $d$ as a function of the energy density for
thermal (solid lines) and atypical states (dashed lines). 
}\label{fig:doubleOccCurves}
\end{figure}

In Fig.~\ref{fig:doubleOccCurves} we present results for the double
occupancy as a function of the energy density for thermal states at
several values of the interaction strength $u$. These are compared to
the corresponding results for the finite entropy density atypical
states with $n=4$ constructed in Section~\ref{app:exactFamilies}.
We see that as the interaction strength $u$ is increased, the results
for thermal and atypical states track one another for an increasing
range of energy densities. On the other hand, for small values of $u$
the double occupancies of thermal and atypical states are very
different at all energy densities. 
\begin{figure}[ht]
\begin{center}
\includegraphics[width=0.45\linewidth]{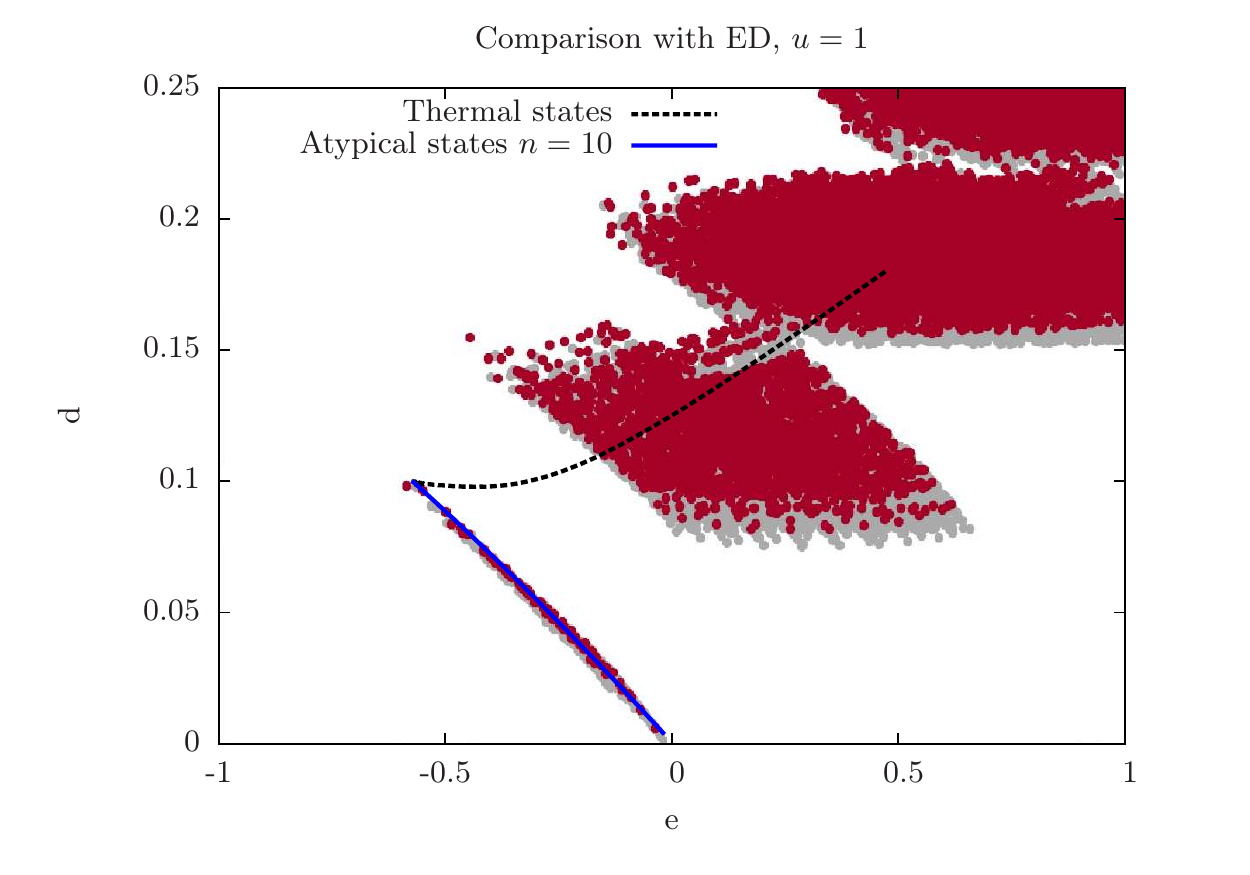}
\includegraphics[width=0.45\linewidth]{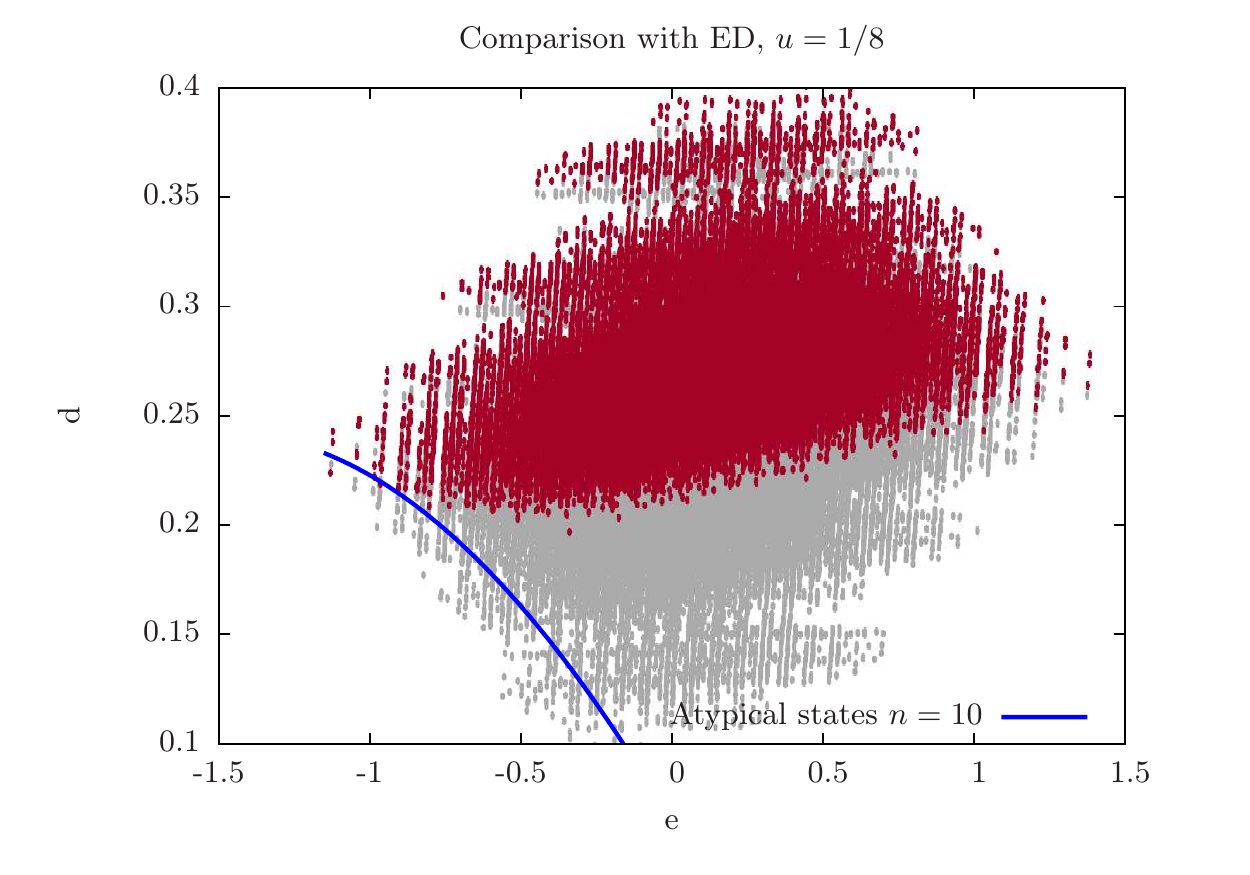}
\end{center}
\caption{
Comparison of thermal and typical states with exact diagonalisation data of Ref.~\onlinecite{hubb1}.
Here the points indicate expectation values for eigenstates of the Hubbard
model, where the spin singlets are highlighted in red.
}\label{fig:HubbEDComp}
\end{figure}
These results can be used to shed
some light on the role played by finite-size effects in the exact
diagonalisation results of Ref.~\onlinecite{hubb1}. There the double
occupancy was computed on lattices of up to $L=12$ sites and a very
interesting change in behaviour of $e(d)$ was observed as a function
of $u$. As shown in Fig.~\ref{fig:HubbEDComp}, for sufficiently
large values of $u$ the there is a ``band'' of eigenstates in the
$d-e$ plane that minimises $d$ at fixed $e$ and is separated from the
region traced out by the other eigenstates. This is easily understood
in terms of typical and atypical eigenstates. The special band of
states is seen to track the $d(e)$ of our $n=4$ family of atypical
states, while for most of the states $d(e)$ is centered around the
result for typical states in the thermodynamic limit (as the system
size in the numerical study is quite small we expect a significant
spread around the thermodynamic limit result). At small values of $u$
the atypical states are no longer visible in $L=12$ numerical data,
which are now all spread around the thermal thermodynamic limit
result. This discrepancy has its origin in the strength of finite-size
effects, which are more pronounced in the small-$u$ limit.



\section{Particular atypical energy eigenstates: the
  ``Heisenberg sector''} \label{sec:Heis} 
In Ref.~\onlinecite{hubb1} it was suggested that a particular class of
eigenstates of the Hubbard Hamiltonian possess the QDL
 property. These states were identified for
short chains by considering the strong coupling regime $t\ll U$. In
this regime the spectrum breaks up into a sequence of narrow ``bands''
of states, which can be characterised by the expectation value of the
double occupancy number operator. The states of interest constitute
the lowest such band and are adiabatically connected to eigenstates
without any doubly occupied sites in the limit $t/U\to 0$.
As we are interested in the limit $L\to\infty$ at fixed $U$, our first
task is to identify such states in terms of the Bethe Ansatz solution.
This can be done either at the level of micro states in a (large)
finite volume, or in terms of macro states in the thermodynamic limit
\cite{QDLHubbard}. 
\subsection{Micro states}
\label{ssec:micro}
An important property of the exact solution of the Hubbard model is
that it makes it possible to follow the evolution of particular
eigenstates with the interaction parameter $u$. In the framework of
the string hypothesis\cite{book} there is a one-to-one
correspondence between energy eigenstates and solutions to the Bethe
Ansatz equations \fr{BAE}, which, in turn, are uniquely characterised by
sets of (half-odd) integers $I_j$, $J^n_\alpha$, $J{'}^n_\alpha$. Fixing
a particular  set $\{I_j, J^n_\alpha, J{'}^n_\alpha\}$ we may follow
the corresponding solution of \fr{BAE} as a function of $u$. This
allows us to identify the special states considered in
Ref.~\onlinecite{hubb1} as follows. In the limit $U\to\infty$ at fixed
$L$ and half filling, the lowest energy states are obtained by
setting 
\be
M'=0\ ,
\ee
i.e. considering only states that do not contain any $k$-$\Lambda$
strings. This is because the latter contribute ${\cal O}(u)$ to the
total energy, \emph{cf.} \fr{eq:energyFull}. These states are
characterised by the quantum numbers $J^n_\alpha$, which have ranges
\be
|J^n_\alpha|\leq\frac{1}{2}\left(\frac{L}{2}-\sum_{m=1}^\infty
t_{nm}M_m-1\right),\quad \alpha=1,\dots,M_n.
\ee
There is no freedom in choosing the $I_j$: they
are given by 
\be
I_j=\begin{cases}
-\frac{L}{2}+j & \text{if } \sum_mM_m \text{ is even}\\
-\frac{L+1}{2}+j & \text{if } \sum_mM_m \text{ is odd}
\end{cases}
\ ,\quad j=1,\dots L.
\ee
Importantly, the $I_j$ form a completely filled ``Fermi sea'', just as
they do in the ground state of the half-filled Hubbard model.
It follows from the results of Ref.~\onlinecite{completeness} that the
total number of such states is $2^L$. We call these states
\emph{Heisenberg sector states}. 
\subsection{Macro states}
At the level of macro states the Heisenberg sector corresponds to the 
requirement
\be
\rho^h(k)=0={\sigma'}_n^p(\Lambda)\ ,\ n=1,2,\dots
\label{HeisenbergBA}
\ee
We note that the correspondence between \fr{HeisenbergBA} and the
microscopic definition of \ref{ssec:micro} is to be understood in a
thermodynamic fashion. There clearly will be eigenstates that are
captured by \fr{HeisenbergBA}, but go beyond the narrow specification
we used in \ref{ssec:micro}. For example, adding a finite number of
$k$-$\Lambda$ strings will not change the macro state
\fr{HeisenbergBA}, as this affect the densities only to order ${\cal
  O}(L^{-1})$. 

Importantly, the ``freezing'' of the charge degrees of freedom that
characterises the Heisenberg sector implies that the thermodynamic
entropy density for these macro states depends only on the spin
degrees of freedom
\be
s=\sum_{n=1}^\infty\int_{-\infty}^\infty d\Lambda\
{\cal S}\left[\sigma_n^p(\Lambda),\sigma_n^h(\Lambda)\right].
\label{SH}
\ee

\subsubsection{Maximal entropy states in the Heisenberg sector}
The next question we want to address is which macro states in the
Heisenberg sector maximise the entropy at a given energy
density. These states would be selected with probability $1$ if one
randomly picked an eigenstate at a given energy density in an
asymptotically large system. We start from the thermodynamic limit of
the Takahashi equations \fr{densities} for Heisenberg sector states
\begin{eqnarray}
\rho^p(k)&=&\frac{1}{2\pi}+\cos
k\sum_{n=1}^\infty\int_{-\infty}^\infty d\Lambda\ \fa{n}{\l-\sin k}
\sigma_n^p(\Lambda)\ ,\nn
\sigma_n^h(\Lambda)&=&-\sum_{m=1}^\infty
\int_{-\infty}^\infty d\Lambda'
A_{nm}(\Lambda-\Lambda')\ \sigma_m^p(\Lambda')
+\int_{-\pi}^\pi dk\ \fa{n}{\sin k-\Lambda}\ \rho^p(k)\ .
\label{densitiesHS}
\end{eqnarray}
We then define an analogue of the free energy density by
\be
f=e-{\cal T}s\ ,
\label{eff}
\ee
where $e$ and $s$ are the energy and entropy densities of Heisenberg
sector states and are given by \fr{entropy} and \fr{SH}
respectively. The ``temperature'' ${\cal T}$ is understood simply as a
Lagrange parameter that allows us to fix the energy density.  We now
extremise the free energy with respect to the particle and hole
densities, subject to (\ref{densitiesHS}). This fixes the ratios 
$\eta_n(\Lambda)=\frac{\sigma_n^h(\Lambda)}{\sigma^p_n(\Lambda)}$ to
be solutions to the system of TBA-like equations
\be
\ln\big(1+\eta_n(\Lambda)\big) =
\frac{g_1(\Lambda)}{{\cal T}}
+\sum_{m=1}^\infty\int_{-\infty}^\infty d\Lambda'\
A_{nm}(\Lambda-\Lambda')\ln\left[1+\frac{1}{\eta_m(\Lambda')}\right],
\label{TBAHS}
\ee
where $g_1(\Lambda)=-4{\rm     Re}\sqrt{1-(\Lambda-inu)^2}+4nu$.
The entropy density for these macro states is given by
\be
s=\sum_{n=1}^\infty\int_{-\infty}^\infty d\Lambda\left[
\frac{g_1(\Lambda)}{{\cal T}}\sigma_n^p(\Lambda)+
g_2(\Lambda)\ln\big(1+\eta_n^{-1}(\Lambda)\big)
\right],
\ee
where $g_2(\Lambda)=\frac{1}{\pi}{\rm Re}\frac{1}{\sqrt{1-(\Lambda+inu)^2}}$.

\section{Entanglement entropy of Heisenberg sector states}
\label{sec:five}
As we have seen above, the Bethe Ansatz solution of the Hubbard model
provides us with a means to compute the thermodynamic entropy density
for any macro state. On the other hand, the notion of a QDL involves
entanglement properties after a partial measurement. Implementing such
partial measurements in the Bethe Ansatz framework is beyond the currently
available methods. However, some information about entanglement
properties of energy eigenstates can be inferred as follows. For
short-ranged Hamiltonians there is a relation between the
thermodynamic and entanglement entropies: if we consider a large
subsystem $A$ of size $|A|$ in the thermodynamic limit, the volume
term in the EE of an eigenstate $|\Psi\rangle$ is
given by 
\be
S_{\rm vN,A}=s|A|+o(|A|),
\label{SA}
\ee
where $s$ is the thermodynamic entropy density. As we have seen
in \fr{SH}, the thermodynamic entropy density of Heisenberg sector
states depends only on the spin degrees of freedom. This then implies that
the volume term in the EE is independent of the
charge degrees of freedom, and depends on the spin sector only. In
particular, as \fr{SA} is based only on the properties of the macro
state under consideration, we know that microscopic rearrangements in
the charge sector, such as introducing $k$-$\Lambda$ strings, will not
affect \fr{SA}. The emerging picture is consistent with expectations
for a QDL state: the spin degrees of freedom exhibit a volume law 
 EE, while the charge degrees of freedom are only
weakly entangled. The spin degrees of freedom will be ``heavy'' in the
terminology of Ref.~\onlinecite{qdl} at large values of $U$, because
their bandwidth is proportional to $t^2/U$. The bandwidth of the
charge degrees of freedom remains ${\cal O}(t)$ and they are therefore
``light'' in comparison. We stress that Heisenberg states obey \fr{SA}
for any value of $u$, and the heavy vs light separation is not
required. This is presumably a consequence of integrability.

Considerations based on the von Neumann EE fall
short of the full QDL diagnostic proposed in Ref.~\onlinecite{qdl},
which requires carrying out a partial measurement of the spin degrees
of freedom. Evidence based on a strong coupling analysis that supports
the view that Heisenberg sector states pass the full QDL diagnostic
has been put forward in Ref.~\onlinecite{QDLHubbard}.
\section{Thermal states in the large-$U$ limit} \label{sec:thermal}
We have constructed an exponential (in the system size) number of
eigenstates, which exhibit QDL behaviour in the thermodynamic limit.
However, these states are atypical in the sense introduced above: the
most likely states at a given energy density are thermal. It is
therefore instructive to contrast the entanglement properties of
Heisenberg sector states with those of typical states. The latter are
given as solutions of the systems \fr{TBAeqns} and \fr{densities} of
coupled integral equations. While it is not possible to solve
these analytically in general, in the limit of strong interactions
analytic results can be obtained \cite{takahashiTBA2,Ha,EEG}. This is
also the most interesting in the QDL context, as it provides a natural
notion of light (charge) and heavy (spin) degrees of freedom. 
It is instructive to 
focus on the ``spin-disordered regime''  
\be
\frac{4t^2}{U}\ll T\ll U\ . \label{eq:spinDis}
\ee
This corresponds to temperatures that are small compared to
the Mott gap, but large compared to the exchange energy.
This is a natural regime in which one may expect the proposed physics
to be realised. Here one has\cite{EEG}  
\be
\rho^h(k)={\cal O}\big(e^{-u/T}\big)\ ,\quad
{\sigma'}^{p,h}_n(\Lambda)={\cal O}\big(e^{-u/T}\big) .
\ee
Substituting this into the general expression \fr{entropy} for the
thermodynamic entropy density we obtain
\be
s=\sum_{n=1}^\infty\int_{-\infty}^\infty d\Lambda\
{\cal S}\left[{\sigma_n}^p(\Lambda),{\sigma_n}^h(\Lambda)\right]
+{\cal O}\Big(\frac{u}{T}e^{-u/T}\Big).
\label{ST}
\ee
Finally, using the relation between thermodynamic and EE
\fr{SA} we conclude that for thermal states in the
spin-disordered regime the contribution of the charge degrees of
freedom contribute to the volume term is
\bea
S_{\rm vN,A}&=&\big(s_{\rm spin}+s_{\rm
  charge}\big)|A|+o\big(|A|\big)\ ,\nn
s_{\rm spin}&=&{\cal O}(1)\ ,\quad
s_{\rm charge}={\cal O}\Big(\frac{u}{T}e^{-u/T}\Big)\ .
\label{weak}
\eea
Here $s_{\rm charge}$ includes the contributions from pure charge
degrees of freedom as well as bound states of spin and charge.
Importantly, unlike Heisenberg sector states, typical states have a
contribution from the charge degrees of freedom to the volume
term. However, this contribution is exponentially small in $u/T$ and
therefore only visible for extremely large subsystems. While we have
focused on the spin-disordered regime, the behaviour \fr{weak}
extends to thermal states for all $0<T\ll U$. In
Ref.~\onlinecite{QDLHubbard}, behaviour of the kind \fr{weak} (for the
entanglement entropy after a partial measurement) was proposed as
a ``weak'' form of a QDL, which one may expect to occur quite
generically in strong coupling limits.

\section{Conclusions}
We have presented evidence that the QDL state of matter
is realised in the strong sense proposed in Ref.~\onlinecite{qdl} for a
particular class of eigenstates of the one-dimensional half-filled Hubbard
model. We have  defined the states constituting this class in the framework of
the Bethe Ansatz by freezing the charge degrees of freedom in the configuration
corresponding to the half-filled ground state. 
For such states, we have explicity shown that the charge degrees of freedom
(``light particles'') do not contribute to the volume term of the bipartite EE.

The Heisenberg sector states are unusual in a precise sense: the EE for typical
(thermal) states at a finite energy density has a volume-law contribution that
involves the charge degrees of freedom even for large $U/t$. We have
shown that in a particular regime at strong coupling the contribution is of the
form \fr{weak}, i.e. 
\beA
S_{\rm vN,A}&=\big(s_{\rm spin}+s_{\rm
  charge}\big)|A|+o\big(|A|\big),\\
s_{\rm spin}&={\cal O}(1)\ ,\quad
s_{\rm charge}={\cal O}\Big(\frac{u}{T}e^{-u/T}\Big)\ ,\quad
\eeA
and we have argued that for $U\gg t$ this form is obtained quite generally at
energy densities that are small compared to $U$. We expect a similar
volume law to occur in the EE after a measurement of all spins. This
suggests the notion of a ``weak'' variant of a QDL, which is
characterised a volume term in the EE after measurement of the heavy
degrees of freedom that is parametrically small, and practically
unobservable except in extremely large systems. We expect that this
weak scenario is not tied to integrability, and should be realised
quite generally in strong coupling regimes.  This is supported by the
strong-coupling expansion results of Ref.~\onlinecite{QDLHubbard}, which do
not rely on integrability.

Given that Heisenberg sector states are atypical and hence rare, a
natural question is how they can be accessed in practice. In principle
this can be achieved by means of a quantum quench \cite{EFreview} from
a suitably chosen initial state. As the Hubbard model is integrable
the expectation values of its infinite number of conservation laws are
fixed by the initial state. At late times after the quench the system
relaxes locally to an atypical state that is characterised by these
expectation values. This provides a general mechanism for realising atypical
states. In order to access a Heisenberg sector state the initial
conditions need to be fine tuned. It would be interesting to
investigate what class of initial states will give rise to Heisenberg
sector steady states. Realising the weak variant \fr{weak} of a QDL is a much
simpler matter. Fixing the energy density to be small compared to the
Mott gap is sufficient to obtain a non-equilibrium steady state that
realises the weak form of a QDL. In the Hubbard model this corresponds
to the spin-incoherent Mott insulating regime. Dynamical properties in
this regime can be analyzed by numerical as well as strong-coupling
methods\cite{spinincoherent}. 

\acknowledgments
We are grateful to P. Fendley, J. Garrison, T. Grover, L. Motrunich and M.
Zaletel for helpful discussions. This work was supported by the EPSRC under
grant EP/N01930X (FHLE), by the National Science Foundation,
under Grant No. DMR-14-04230 (MPAF) and by the Caltech Institute of
Quantum Information and Matter, an NSF Physics Frontiers Center with
support of the Gordon and Betty Moore Foundation (MPAF).

\end{document}